\documentclass[twocolumn,showpacs,preprintnumbers,amsmath,amssymb]{revtex4}
\usepackage{graphics}
\usepackage{epsfig}
\usepackage{bm}

\begin{document}
\title{Cooling and heating by adiabatic magnetization in the Ni$_{50}$Mn$_{34}$In$_{16}$ magnetic shape memory alloy}

\author{Xavier Moya, Llu\'{i}s Ma\~nosa}
\email{lluis@ecm.ub.es}
\author{Antoni Planes}

\affiliation{Departament d'Estructura i Constituents de la
Mat\`eria, Facultat de F\'isica, Universitat de Barcelona,
Diagonal 647, E-08028 Barcelona, Catalonia, Spain}

\author{Seda Aksoy, Mehmet Acet, Eberhard F. Wassermann}
\author{Thorsten Krenke}
\altaffiliation{Present address: ThyssenKrupp Electrical Steel
GmbH, D-45881 Gelsenkirchen, Germany} \affiliation{Fachbereich
Physik, Experimentalphysik, Universit\"{a}t Duisburg-Essen,
D-47048 Duisburg, Germany}

\date{\today}

\begin{abstract}

We report on measurements of the adiabatic temperature change in
the inverse magnetocaloric Ni$_{50}$Mn$_{34}$In$_{16}$ alloy. It
is shown that this alloy heats up with the application of a
magnetic field around the Curie point due to the conventional
magnetocaloric effect. In contrast, the inverse magnetocaloric
effect associated with the martensitic transition results in the
unusual decrease of temperature by adiabatic magnetization. We
also provide magnetization and specific heat data which enable to
compare the measured temperature changes to the values indirectly
computed from thermodynamic relationships. Good agreement is
obtained for the conventional effect at the second-order
paramagnetic-ferromagnetic phase transition. However, at the first
order structural transition the measured values at high fields are
lower than the computed ones. Irreversible thermodynamics
arguments are given to show that such a discrepancy is due to the
irreversibility of the first-order martensitic transition.
\end{abstract}


\maketitle


\section{Introduction}

When the magnetization of any magnetic material is changed
isothermally under the application of a magnetic field, heat is
exchanged with the surroundings. If the change is performed
adiabatically, the temperature changes. This is the magnetocaloric
effect (MCE), which provides the basis of the adiabatic
demagnetization cooling technique \cite{Tishin2003}. This
technique was developed to reach mK temperatures soon after the
pioneering work by Debye \cite{Debye1926} and Giauque
\cite{Giauque1927}, who independently suggested such a possibility.
The discovery in the nineties of the giant magnetocaloric effect
associated with first-order magnetostructural transitions in a
number of intermetallic alloy families \cite{Pecharsky1997}
opened up the possibility of using this technique in room
temperature refrigeration applications and, thus, yielded renewed
interest in the subject \cite{Bruck2005}.

It has been known for a long time that the isothermal reduction of
a magnetic field gives rise to a decrease in entropy  in some
antiferromagnetic and ferrimagnetic systems,
\cite{Joenk1962,Clark1969}. This inverse magnetocaloric phenomenon
was supposed to produce small effects and has been largely
ignored. Recently, however, it has been shown that in some
ferromagnetic \cite{Krenke2005} and metamagnetic
\cite{Sandeman2006} systems, inverse MCE can have an amplitude
comparable to the conventional effect detected in giant
magnetocaloric intermetallic materials. The inverse effect is
related to the existence of regions in phase space where
$\zeta =(\partial M /\partial T)_H$ is positive. In a paramagnetic
system, $\zeta$ is always negative, and thus, the origin of a
positive $\zeta$ must be ascribed to coupling between magnetic
moments. The inverse MCE can occur in the vicinity of magnetostructural and
metamagnetic phase transitions due to changes in the magnetic
coupling driven by the interplay between magnetic and structural
degrees of freedom \cite{book2005}.

In the present paper, we study the MCE in a
Ni$_{50}$Mn$_{34}$In$_{16}$ alloy. This is a magnetic shape-memory
alloy which undergoes a martensitic transition from a cubic
($L2_1$) to a monoclinic ($10M$) structure below its Curie
temperature \cite{Krenke2006a}. Interestingly, the sample shows
both inverse and conventional MCE in rather close temperature
intervals. While the conventional effect arises from the
continuous transition from paramagnetic to ferromagnetic states,
the inverse effect is associated with the martensitic transition
at which the magnetic moment of the system decreases. This
decrease originates from the tendency of the excess of Mn atoms
(with respect to 2-1-1 stoichiometry) to introduce
antiferromagnetic coupling. The antiferromagnetic coupling is
caused by the change in the Mn-Mn distance as the martensitic
phase of lower symmetry gains stability \cite{Brown2006}.

While most of the reported data on giant MCE materials refer to
the isothermal entropy change, the most relevant parameter for
actual applications of this effect is the adiabatic temperature
change \cite{Pecharsky2006}. This value is usually computed from
entropy data by means of equilibrium thermodynamic relationships.
However, irreversible effects are expected to take place at
first-order phase transitions which can yield discrepancies
between the computed temperature change and the directly measured
one. Actually, direct measurements of the temperature change in
giant MCE compounds are scarce, and the reported values in many
cases do not seem to be consistent with those indirectly computed
\cite{Giguere1999,Gschneider2000,Pasquale2005}. Here, we report on
adiabatic temperature measurements, which provide direct evidence
of cooling by adiabatic magnetization in an inverse magnetocaloric
material. It is also shown that heating is achieved at the
paramagnetic-ferromagnetic phase transition. We focus on moderate
magnetic fields which are readily available for applications of
giant MCE materials \cite{Pecharsky2006}. Furthermore, data
obtained from magnetization and heat capacity experiments have
enabled us to compare the measured temperature change with that
computed from entropy data. Irreversible thermodynamics arguments
are provided to account for the discrepancies observed at the
first-order structural phase transition.

\section{Experimental details}

A polycrystalline Ni$_{50}$Mn$_{34}$In$_{16}$ ingot was prepared
by arc melting the pure metals under argon atmosphere in a
water-cooled Cu crucible and subsequently re-melted in order to
ensure homogeneity. The ingot was sealed under
argon in a quartz recipient and annealed at 1073 K for 2 hours.
Finally, it was quenched in ice-water. The composition of the
alloy was determined by energy dispersive X-ray photoluminescence analysis
(EDX). For calorimetric and magnetization measurements, a small
sample (61.5 mg)  was cut using a low-speed diamond saw. The
remaining button (13 mm in diameter, 6 mm thickness and 4.6
g) was used for the adiabatic temperature change measurements.

Magnetization was measured by means of a SQUID magnetometer, and
differential scanning calorimetric (DSC) measurements were
conducted using a high-sensitivity calorimeter. Specific heat
measurements were performed using a modulated differential
scanning calorimeter (MDSC), and data were taken with the constant
temperature method \cite{Boller94} starting from the lowest
temperature (190 K).

Adiabatic temperature changes were measured at atmospheric
pressure using a specially designed set-up. A thin (0.75 mm
diameter) Ni-Cr/Ni-Al thermocouple was used to measure the
temperature. The output of this thermocouple was continuously
monitored by means of a multimeter that also electronically
compensates for the reference junction. Measurements without any
specimen confirmed that the recorded values were not affected by
magnetic fields up to 1.3 T. The thermocouple was embedded within
the sample and good thermal contact between the sample and the
thermocouple was ensured by Ariston conductive paste. The sample
is situated inside a copper container (sample holder), which is
placed on the top face of a Peltier element. The bottom surface
sits on a copper cylinder, which acts as a heat sink. The bottom
end of the cylinder is in contact with a nitrogen bath. By
controlling the current input into the Peltier element, it is
possible to achieve fine tuning of the temperature in the 200-320
K interval. Temperature oscillations were less than 0.05 K.
Thermal insulation (adiabaticity) between the sample and sample
holder was ensured by a polystyrene layer. The sample holder was
placed in between the poles of an electromagnet (28 mm gap), which
enabled fields up to 1.3 T to be applied. A major advantage of
using an electromagnet is the short rising time in the application
of the field (the field rises from 0 to 1 T in about 0.5 s). Such
a field rise time is several orders of magnitude shorter than the
thermal relaxation time of the sample-holder system ($\sim$ 100
s), thus ensuring the adiabaticity of the process.

In order to check the reliability of the device, we measured the
MCE of  commercial pure (99.9 wt \%) Gd. The measured temperature
changes obtained around the Curie point for a magnetic field of 1T
agree with those reported in the literature \cite{Dankov98}.

\section{Results and Discussion}

\begin{figure}
\includegraphics[clip=]{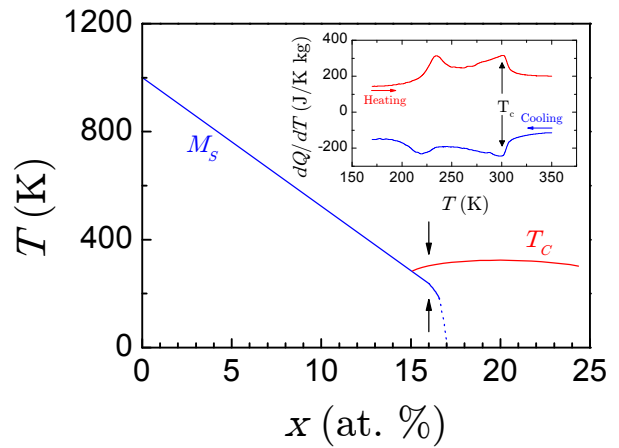}
\caption{(Color online) Phase diagram of
Ni$_{50}$Mn$_{50-x}$In$_{x}$, obtained using the data in ref.
\cite{Krenke2006a}. $M_{S}$ indicates the martensitic transition
line and $T_{C}$ indicates the Curie point line.The inset shows
DSC curves for heating and cooling runs for the $x=16$ sample.}
\label{fig1}
\end{figure}

For the present study we selected a composition with the
para-ferromagnetic and martensitic transition temperatures close
to each other. This is illustrated in figure \ref{fig1}, which
shows the Curie and  martensitic transition start temperatures as
a function of In content for Ni$_{50}$Mn$_{50-x}$In$_x$ alloys.
Continuous lines are polynomial fits to the data given in ref.
\cite{Krenke2006a}. The arrows indicate the composition of the
studied sample. The inset presents DSC curves (heating and
cooling) for the present Ni$_{50}$Mn$_{34}$In$_{16}$ alloy
\cite{thermograms}. The peaks at higher temperature correspond to
the Curie point and those at lower temperatures correspond to the
martensitic transition (which occurs with 15 K thermal
hysteresis). Integration of the peaks associated with the
martensitic transition renders  latent heats of -1750 $\pm$ 100
J/kg for the cooling run (forward transition) and 1850 $\pm$ 100
J/kg for the heating run (reverse transition).

\begin{figure}
\includegraphics[clip=]{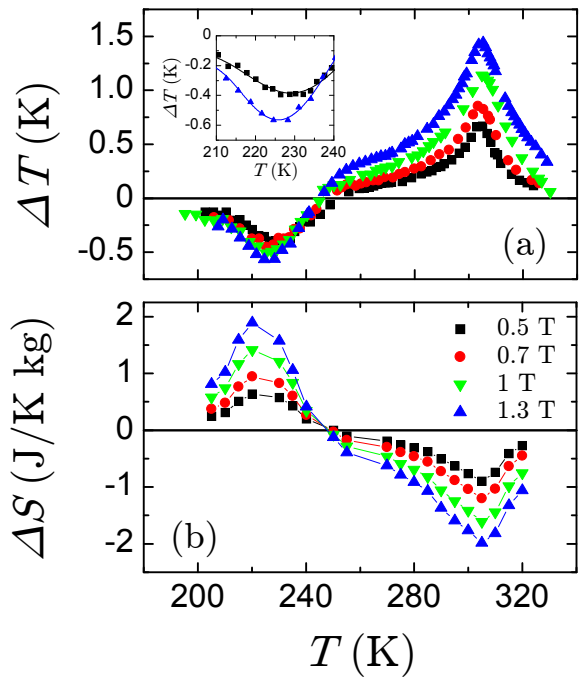}
\caption{(Color online) (a) Measured adiabatic temperature change
and (b) computed isothermal entropy change, as a function of
temperature at selected values of the magnetic field. The inset
shows an enlarged view for the 0.5 and 1.3 T fields which
illustrates the shift in the inverse MCE with magnetic field.}
\label{fig2}
\end{figure}

The adiabatic temperature changes measured over the 200-320 K
temperature range for selected values of the magnetic field are
shown in figure \ref{fig2}(a). Data points were obtained according
to the following procedure, which ensures the suppression of any
history dependent effect: first, the sample is heated up to 320 K
(above the Curie point) and then cooled down to the fully
martensitic state at 170 K. Subsequently, it is heated up to the
desired temperature and the magnetic field is switched on for 20
s. After switching off the field, the sample is heated again above
the Curie point and the protocol is repeated for the next data
point. The measured adiabatic temperature changes shown in fig.
\ref{fig2}(a) prove unambiguously that the sample cools down upon
adiabatic application of the field in the temperature range
200-245 K, while it heats up in the temperature range 245-320 K.
The positive temperature change has its maximum value ($\Delta T
\simeq $1.5 K for 1.3 T) at the Curie point. The maximum
temperature decrease ($\Delta T \simeq$ -- 0.6 K for 1.3 T) occurs
at a temperature that shifts with magnetic field [see inset in
figure \ref{fig2}(a)], in agreement with the decrease in the
martensitic transition temperature reported for Ni-Mn-In alloys
\cite{Krenke2006a}. The values found for $\Delta T$ at their
corresponding peak temperatures are comparable to those reported
for other giant MCE materials. However, a novel feature for
Ni$_{50}$Mn$_{34}$In$_{16}$ is that these relatively large
temperature changes can be either positive or negative.

\begin{figure}
\includegraphics[clip=]{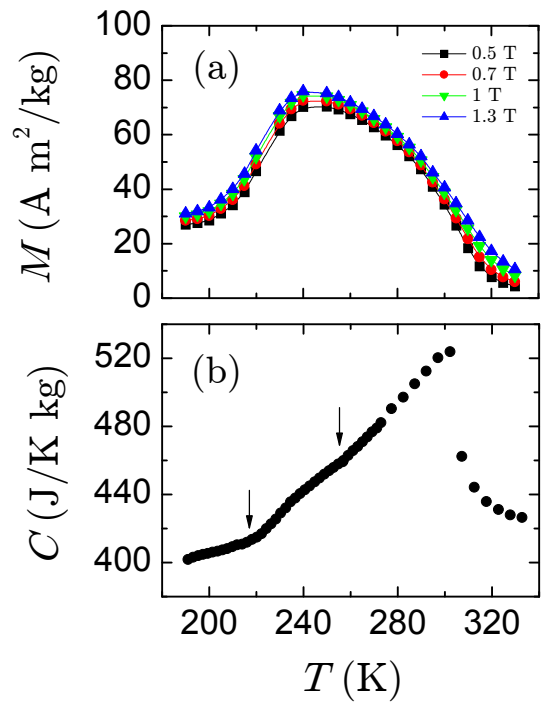}
\caption{(Color online) (a) Temperature dependence of the
magnetization for selected values of the magnetic field. (b)
Specific heat as a function of temperature. Arrows indicate the
region of the reverse martensitic transition.} \label{fig3}
\end{figure}
\begin{figure}
\includegraphics[clip=]{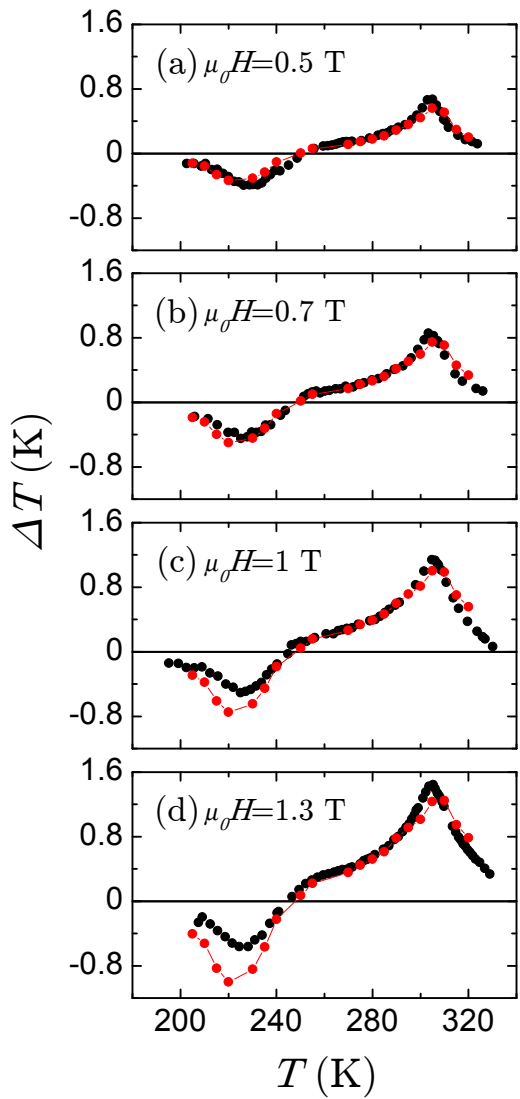}
\caption{(Color online) Adiabatic temperature change, as a
function of temperature, for different magnetic fields. Black
symbols stand for measured data and red symbols correspond to data
indirectly computed using equilibrium thermodynamics
relationships.} \label{fig4}
\end{figure}

In order to correlate the measured temperature changes with those
indirectly computed from entropy data, we measured the
magnetization of the sample as a function of temperature and
magnetic field. Results at selected fields are shown in fig
\ref{fig3}(a). In the temperature range 245-320 K, $\zeta$ is
negative, while a positive $\zeta$ is obtained in the range
200-245 K. From these data, we computed the magnetic field-induced
entropy change by using the Maxwell relation $\Delta S = \mu_{0}
\int  \zeta  dH$. Results are shown in Fig. \ref{fig2}(b).
Excellent qualitative agreement is observed between the two
quantities ($\Delta T$ and $\Delta S$) characterizing giant MCE.
Conventional MCE is observed within the 245-320 K temperature
range, i.e. a negative entropy change with the associated positive
temperature change, while in the 200-245 K interval, the sample
exhibits inverse MCE: an increase in entropy with the associated
negative temperature change.

It is customary to compute the adiabatic temperature change from
isothermal entropy data by means of the following relationship,

\begin{equation}
\Delta T_{rev} = - \frac{T}{C} \Delta S \label{eqn1},
\end{equation}

\noindent which is expected to be valid in equilibrium. $C$ is the
specific heat at constant magnetic field and is assumed to be
independent of the magnetic field. In order to check the validity
of this approach, we measured the specific heat of our
Ni$_{50}$Mn$_{34}$In$_{16}$ sample. Results are shown in figure
\ref{fig3}(b). The large lambda-type peak at 302 K corresponds to
the second-order para-ferromagnetic phase transition. In the
temperature range 216-257 K a small bump is observed, which
coincides with the reverse martensitic transition. No latent heat
contributions are expected for the isothermal-modulated method we
have used.

In Fig. \ref{fig4}, we compare the measured adiabatic temperature
changes (black symbols) with those computed from the entropy [Fig.
\ref{fig2}(b)] and specific heat data [Fig. \ref{fig3}(b)] (red
symbols) for different values of the applied field. Good agreement
between measured and computed values over the complete temperature
range is obtained at low magnetic fields. As the magnetic field is
increased, there is still good agreement between the data
corresponding to conventional MCE \cite{discrepancies}, but the
absolute value of the measured temperature change becomes smaller
than the computed one in the inverse MCE region.  Such a
difference is due to the irreversibility associated with the
first-order phase transition.

In order to consider the effect of dissipation, we start from the
Clausius inequality $\oint \frac{\delta q}{T} \leq 0$, which can
be expressed as $\frac{\delta q}{T} = dS - \delta S_i$, where $dS$
is a reversible differential change of entropy and $\delta S_i$ is
the entropy production ($\delta S_i \geq 0$). When the magnetic
field is adiabatically changed, $\delta q =0$, and under the
assumption of a quasistatic, continuous process with hysteresis
\cite{Ortin2005}, the adiabatic temperature change is expressed
as

\begin{equation}
\Delta T =  \frac{T}{C} \left[ -\Delta S + S_i \right] = \Delta
T_{rev} + \frac{T S_i}{C} \label{eqn2},
\end{equation}

\noindent where $TS_i$ is the dissipated energy ($E_{diss}$). For
an inverse magnetocaloric effect, there is an increase of entropy
by the application of the field, i.e. $\Delta S>0$. On the other
hand, $S_i$ is always positive. Hence, for an out-of-equilibrium
process, the two terms within brackets in equation \ref{eqn2} will
partially cancel each other when the field is swept from zero to a
given value, and therefore, the measured temperature change will
always be less than the value computed using equilibrium
thermodynamics (see equation \ref{eqn1}). Such a difference is
expected to be small at low fields (close to equilibrium
conditions), but it becomes larger at higher fields. Note that for
conventional MCE, when the field changes from 0 to H, $\Delta T
\geq \Delta T_{rev}$, which is consistent with the data around the
Curie point.

At each temperature, the dissipated energy is given by $E_{diss} =
T \Delta S  + C \Delta T$. A value of 158 J/kg is found at 225 K
for a field of 1.3 T. This value amounts to about 10 \% of the
latent heat of the martensitic transition in this alloy.

In giant magnetocaloric materials for which the MCE is associated
with a first-order transition, the giant effect relies on the
possibility of inducing the phase transition by application of a
magnetic field. The martensitic transition is driven by phonon
instabilities in the transverse TA$_2$ phonon branch ($[110]$
propagation and $[1\bar{1}0]$ polarization)
\cite{Planes2001,Entel2006}. Recent {\it ab-initio} calculations
for cubic Ni$_2$MnIn have shown that increasing the magnetization
due to an external field favors the cubic structure and leads to a
gradual vanishing of the phonon instability \cite{Entel2007} due
to the coupling between vibrational and magnetic degrees of
freedom. This effect results in a marked decrease of the
martensitic transition temperature with increasing field that
enables to induce the transition by the application of a field at
a temperature close to the zero field transition temperature.
Hence, the microscopic origin of the inverse MCE in
Ni$_{50}$Mn$_{34}$In$_{16}$ must be ascribed to such
magnetoelastic coupling responsible for the change in the relative
stability of the martensitic and cubic phases.

\section{Conclusion}

By directly measuring the adiabatic temperature change in the
Ni$_{50}$Mn$_{34}$In$_{16}$ alloy, we provide experimental
evidences of both cooling and heating in a giant inverse
magnetocaloric compound. It has been shown that the
irreversibility associated with the first-order structural
transition gives rise to measured temperature changes which are
lower than those indirectly computed using equilibrium
thermodynamics. The existence of a temperature region where the
magnetocaloric effect reverses sign under weakly applied magnetic
fields opens up the possibility of new applications of this
fascinating property.

\begin{acknowledgments}

This work received financial support from the CICyT (Spain),
Project No. MAT2004--01291,  DURSI (Catalonia), Project No.
2005SGR00969, and from the Deutsche Forschungsgemeinschaft
(GK277). XM acknowledges support from DGICyT (Spain). We thank
Peter Hinkel for technical support.
\end{acknowledgments}


\bibliography{apssamp}

\begin{thebibliography}{99}
\bibitem{discrepancies} The small discrepancies in the data around
the Curie point arise from the fact that the specific heat has
been considered to be independent of magnetic field. For
ferromagnetic materials the effect of magnetic field is to smooth
out the peak in the specific heat.
\end{thebibliography}

\end{document}